# Toward a 2D SPH multiphysic code with solid-solid & fluid interactions for industrial related problems

D. Cébron [(a)], J-F. Sigrist [(b)]

[(a)] *École Centrale de Nantes, Nantes 44000, France*

[(b)] *Service Technique et Scientifique, DCNS Propulsion, La Montagne 44620, France*

ABSTRACT: This paper gathers basic principles and literature suggestions on the SPH method from an engineering point of view and for multiphysic applications. This practical review is done through the presentation of a code in which these features have been developed. Applications of the method to industrial related issues are considered by starting from an existing open-source 2D SPH code, namely the SPHYSICS code, which offers an effective ground for numerical developments. Firstly, the different features added to obtain an operational code needed for engineering applications are described, offering a kind of review of SPH methods for engineers. Secondly, the validation of the proposed code is partially presented with two test cases. Thirdly, principles of a method to solve solid/solid contacts, frequently present in realistic configuration, are exposed and applied to achieve more complex simulations.

## 1 INTRODUCTION

Impact of a structure on a fluid is an academic problem of major interest in naval shipbuilding since it is a representative case of the so-called "slamming" situation, which occurs for a surface ship in various operational conditions (see illustration provided by Fig. 1 in the case of a frigate sailing at high speed).

The slamming problem has therefore been extensively studied over the past years, both from the experimental as well as from the numerical points of view, see Donguy (2001) and Peseux et *al*. (2005), among many others on the matter.

As far as numerical techniques are concerned, various approaches can be employed to tackle the impact problem. Some industrial finite codes, such as the LS D-DYNA explicit code, offer some functionality to deal with the slamming problem, see Fig. 2; the underlying methodology is based on the so-called Arbitrary-Lagrangian-Eulerian method with the Multi-Material formalism (Aquelet, 2004; Aquelet et *al*., 2005). Such methodology is rather demanding in terms of modelling and computing effort, and can thus not be resorted to in pre-design analyses using numerical techniques. Smoothed Particle Hydrodynamic (SPH)-based techniques also provide some applications to the slamming problem, see Fig. 3 bellow, extracted from Cébron (2008).

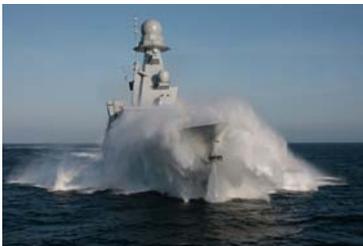
Figure 1. Horizon Frigate at high speed

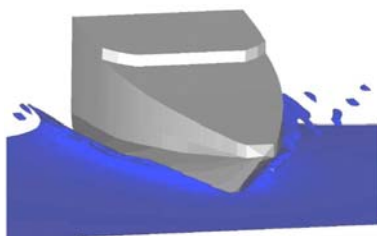
Figure 2. Slamming problem with an industrial FEM code

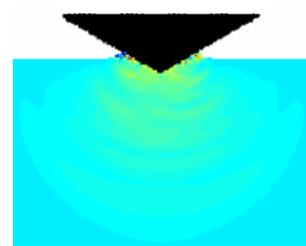
Figure 3. SPH acoustic wave generated by a wedge impact

The present paper aims at investigating various applications of the SPH approach for problems involving Multi-Fluid & Solid systems (MF&S), with the view to spreading the use of SPH techniques for industrial applications.

Adopting an engineering point of view, we have undertaken some numerical developments within an existing SPH-based free license code, namely the SPHYSICS code (Dalrymple-Rogers, 2007;

Crespo et al., 2007) in order to perform some applications in MF&S situations. A contribution to some enhancement of the SPHYSICS code is exposed and discussed in the present paper. The work we have undertaken leads to a potential up-dated SPHYSICS[2] version of the original code, which, far from being fully operational for the time being, yields however encouraging results and identifies promising developments.

The paper is organized as follows: in the first subsection, the main features of the SYPHYSIC[2] code are presented, starting from the existing functionalities of the SPHYSIC code, which are also briefly recalled; in the second subsection, elementary test cases are studied, namely the dam break problem and the wedge impact problem, which allows some validation of the SYPHYSIC[2] code; in the third subsection, extension of the SYPHYSIC[2] code toward problems involving solid/solid and solid/fluid interactions are discussed.

## 2 FEATURES OF SPHYSICS[2]

### 2.1 *Context and background of the SPHYSICS[2]*

SPHYSICS is a SPH code inspired by the formulation of Monaghan (1992) and developed jointly by researchers at the Johns Hopkins University (USA), the University of Vigo (Spain), the University of Manchester (UK) and the University of Rome La Sapienza (Italy). SPHYSICS is a free license code and has been validated in many configurations, as described by Dalrymple & Rogers (2007), Crespo et al. (2007) and Gomez et al. (2007). In such context, additional features of the SPH-based core of the program have been developed, in order to tackle more and more complex coupled problems on the one hand, and to spread the use of SPH techniques for industrial pre-design studies, on the other hand. Since the SPH method is well described in many papers (Monaghan, 1992; Monaghan, 2005; Benz, 1989), only some basic principles of the method are recalled in the present paper; they are supplied with SPHYSICS and are provided here for the shake of basic comprehension. The SPH method is based on integral interpolants, which approximate any function $f_{(\vec{r})}$ according to the relation:

$$f_{(\vec{r})} = \int f_{(\vec{r}')} \cdot W_{(\vec{r}-\vec{r}',h)} d\vec{r}' \qquad (1)$$

where $h$ is called the "smoothing length" and $W_{(\vec{r}-\vec{r}',h)}$ is the "weighting function", sometimes referred to as "kernel function". In a discrete form, Eq. (1) leads to the following approximation of the function at a particle $I$ (or "interpolation point"):

$$f_i = \sum_j m_j \frac{A_j}{\rho_j} W_{ij} \qquad (2)$$

where the summation is over all the particles within the region of compact support of the kernel function. The mass and density are denoted by $m_j$ and $\rho_j$ respectively and $W_{ij} = W_{(\vec{r}_i-\vec{r}_j,h)}$ is the kernel.

The performance of an SPH model is critically dependent on the choice of the weighting functions. They should satisfy several conditions such as positivity, compact support, and normalization. Also, $W_{ij}$ must be monotonically decreasing with increasing distance from particle $i$ and behave like a delta function as the smoothing length $h$ tends to zero. The code SPHYSICS proposes different types of kernel functions: in our developments, we therefore have not worked on this aspect of the SPH method.

### 2.2 *Dynamic equations and various improvements*

The conservation of the momentum used in SPHYSICS is the Euler equation, which is

$$\frac{D\vec{v}}{Dt} = -\frac{\nabla p}{\rho} + \vec{\Theta} + \frac{1}{\rho}\vec{F} \qquad (3)$$

where $\vec{\Theta}$ refers to the diffusion terms, such as $\vec{\Theta} = \nu \cdot \nabla^2 \vec{v}$ in the Navier-Stokes equation, or the well known artificial viscosity $\Pi$. The force $\vec{F}$ is the sum of the forces acting on the continuum field, e.g. inertial forces, gravitation, etc... It is recalled that in SPHYSICS, the diffusion term can account for three different kind of viscosity: artificial, laminar and full viscosity, including in this latter case laminar viscosity and Sub-Particle Scale (SPS) turbulence. s studied in (Falappi-Gallati, 2007), the diffusion term can also be used to model granular material or ice floes drifting under the action of the wind (Oger-Savage, 1999; Gutfraind-Savage, 1997). Assuming that the granular flow type is mainly frictional, the following law, which includes a Mohr-Coulomb yield criterion, allows reproducing the friction between grains and the stopping of the granular landslide:

$$\vec{\Theta} = \frac{1}{\rho}(\nabla \cdot \eta_{eff} \nabla) \vec{v} \quad (4)$$

with the effective viscosity $\eta_{eff} = \min\left[\frac{p \sin(\varphi)}{\sqrt{4I_{2S}}}, \eta_{max}\right]$ supposed to be variable in space, $\varphi$ the internal friction angle and $I_{2S}$ the second invariant of the rate of strain tensor. Another expression with three constitutive constants, namely $\mu_s$, $\mu_2$ and $I_0$, is suggested in Pouliquen et al., (2006):

$\eta_{eff} = \frac{p}{\sqrt{4I_{2S}}}\left[\mu_s + \frac{\mu_2 - \mu_s}{1 + I_0/I}\right]$, where $I = d\sqrt{\frac{I_{2S}}{P/\rho_P}}$ for particles of diameter $d$ and density $\rho_p$.

The Euler equation, is a simplified form of the more general equation formulated bellow:

$$\frac{D\vec{v}}{Dt} = \frac{1}{\rho}\text{div}(\overline{\overline{\sigma}}) + \frac{1}{\rho}\vec{F} \quad (5)$$

In order to take into account different rheological laws for the stress tensor $\overline{\overline{\sigma}}$, the Euler equation has been replaced by Eq. (4) in SPHYSICS[2]. Following [10], $\overline{\overline{\sigma}}$ is calculated as $\sigma^{ij} = -p \cdot \delta^{ij} + S^{ij}$ where $-p \cdot \delta^{ij}$ is the spherical part of $\overline{\overline{\sigma}}$, and $S^{ij}$ is the deviatoric stress. In linear elastic theory, $S^{ij}$ can be obtained from Hooke's law: with $\mu$ the shear modulus,

$$\frac{dS^{ij}}{dt} = 2\mu\left(\dot{\varepsilon}^{ij} - \frac{1}{3}\delta^{ij}\dot{\varepsilon}^{kk}\right) + S^{ik}R^{jk} + R^{ik}S^{kj} \quad (6)$$

In Eq. (6), $\overline{\overline{\dot{\varepsilon}}}$ is the deformation tensor and $\overline{\overline{R}}$ is the rate of rotation tensor defined by [10]:

$$\overline{\overline{\dot{\varepsilon}}} = \frac{1}{2}(\nabla\vec{v} + {}^t\nabla\vec{v}) \quad (7)$$

$$\overline{\overline{R}} = \frac{1}{2}(\nabla\vec{v} - {}^t\nabla\vec{v}) \quad (8)$$

The mass conservation equation imposes:

$$\frac{D\rho}{Dt} + \rho \nabla \cdot \vec{v} = 0 \quad (9)$$

The two previous principal equations – namely Eqs. (3) and (9) – can be written using different SPH formulation; in SPHYSICS[2], a possibility of choosing the formulation has been added, using the general form given below with an artificial diffusion term:

$$\frac{d\vec{x}_a}{dt} = \vec{v}_a \quad (10)$$

$$\frac{d\rho_a}{dt} = \sum_b m_b \left(\frac{\vec{v}_a - \vec{v}_b}{\rho_a^{\sigma-2}\rho_b^{2-\sigma}}\right)\nabla W_{ab} \quad (11)$$

$$\frac{dv_a^i}{dt} = \frac{F^i}{\rho} + \sum_b m_b \left(\frac{\sigma_a^{ij}}{\rho_a^{\sigma-2}\rho_b^{2-\sigma}} + \frac{\sigma_b^{ij}}{\rho_a^{\sigma-2}\rho_b^{2-\sigma}} + \Pi_{ab}\right)\frac{\partial W_{ab}}{\partial x_a^j} \quad (12)$$

which allows recovering the Monaghan (1992) formulation by setting $\sigma = 2$, and the formulation used by Vila (1999) and Colagrossi & Landrini (2003) by setting $\sigma = 1$. As highlighted by Colagrossi & Landrini (2003), this latter formulation is very interesting because it allows multi-phasic flows. In order to deal with multiphase flows[1], the viscosity and divergence operator corrections proposed by Colagrossi & Landrini (2003) have been implemented in the code.

## 2.3 Equations of state for fluid and elastic materials

One of the main SPH features consists in avoiding an expensive resolution of the Poisson equation in considering any material as compressible. It implies the use of an Equation Of State (EOS) for each material in the simulation. For fluids, the usual equation used to describe acoustic wave propagation, originated from adiabatic transformations of an ideal gas, is written:

$$P - P_0 = B\left[\left(\frac{\rho}{\rho_0}\right)^\gamma - 1\right] = \frac{\rho_0 c_0^2}{\gamma}\left[\left(\frac{\rho}{\rho_0}\right)^\gamma - 1\right] \tag{13}$$

In Eq. (13), also referred to as the Tait's equation, $c_0^2 = \left.\frac{dP}{d\rho}\right|_{\rho=\rho_0}$ is the square of nominal value of the chosen sound speed, $\rho_0$ is the nominal density value, and $\gamma$ is the polytropic constant of the fluid ($\gamma \approx 7$ for water, and $\gamma \approx 1.4$ for an ideal gas, according to Tait (1888)). It is interesting to notice that the hydrostatic pressure initialisation has to consider this weak compressibility of the fluid: $\frac{\partial P}{\partial z} = g\rho_{(z)}$ implies that, following Eq. (13), $\rho(z)$ is $\rho = \rho_0\left(1 - \frac{(\gamma-1)g}{c_0^2}z\right)^{\frac{1}{\gamma-1}}$, with $z = 0$ at the free surface; the pressure field can be recovered from the EOS, once $\rho_{(z)}$ is set. As noted by Doring (2005), this latter expression of $\rho$ differs from the one proposed by Monaghan (1994): $\rho = \rho_0\left(1 - \frac{g}{c_0^2}z\right)^{\frac{1}{\gamma}}$. It is a good approximation but is not the analytical solution of the hydrostatic problem.

Even if test cases with elastic material are not considered in the present paper, it is important to note that a solid described through the SPH formalism needs also an EOS, such as the Tillotson or the Mie-Gruneisen equation. In order to study elastic material or dynamical fracture of brittle material (Garai, 2007; Oger et al., 2006), the potential function proposed in Garai (2007) has been used to add a solid equation of state in the code.

With a potential function of the form $U_{(r)} = -\frac{A}{r^m} + \frac{B}{r^n}$, the following solid EOS is derived:

$$P - P_0 = 3\frac{\rho_0 c_0^2}{n-m}\left[\left(\frac{\rho}{\rho_0}\right)^{1+\frac{n}{3}} - \left(\frac{\rho}{\rho_0}\right)^{1+\frac{m}{3}}\right] \tag{14}$$

where $r$ is the interatomic spacing and $A$, $B$, $m$ and $n$ are constants.

As suggested in Colagrassi-Landrini (2003), a cohesion force can be added if necessary. This is done in adding the term $-\bar{a}\cdot\rho^2$ in previous EOSs, where $\bar{a}$ controls the strength of the cohesion force. The pressure gradient has to be written as:

$$\nabla p_i = \sum_j \frac{m_j}{\rho_j}(p_j + p_i)\cdot\nabla W_j(x_i) - \bar{a}\cdot\sum_j \frac{m_j}{\rho_j}(\rho_j^2 + \rho_i^2)\cdot\nabla W_j(x_i) \tag{15}$$

The sound speed is usually arbitrary chosen to have a field Mach number below 0.1 and then the SPH solutions remains close to the incompressible solution (Monoghan, 1994). Indeed, because of stability requirements, the higher the sound speeds in material, the smaller the time step is: this constraint can then lead to numerical problem, when dealing with elastic materials.

---

[1] *The multiphase flow option of our code is not discussed in the present paper; it is however under development and still requires some validations. This will be dealt with in future publications.*

### 2.4 *Local pressure evaluation and free motion*

Many problems in Computational Fluid Dynamic imply to know local pressure on solid boundaries, especially the simulation of the free motion for a rigid body. However, as recalled by Martin & Moyce (1952), this point is usually not addressed in the SPH literature. In fact, the pressure estimation is a subtle work because pressure irregularities remains even if the artificial viscosity term helps obtaining a smoother pressure distribution.

In a first attempt to calculate local pressure, a procedure close to the method proposed by Oger et *al.* (2006) has been implemented in SPHYSICS[2]: instead of improving the pressure distribution, for instance by periodic re-initialisation, a second order accurate interpolation, through a MLS kernel as discussed by Colagrossi & Landrini (2003) has been resorted to for that purpose. After validation of this local pressure evaluation procedure (see next subsection), a coupling fluid/solid procedure has been included in our code, as described by Oger et *al.* (2006). Dealing with fluid/solid systems implies solving the additional rigid body equations $\frac{d\vec{V}_{solid}}{dt} = \frac{1}{M}\vec{F}_{ext \rightarrow solid}$ and $\frac{d^2\theta}{dt^2} = \frac{1}{I_G}M_{ext \rightarrow solid}$ where $M$ and $I_G$ are the mass and the inertia taken in the centre of gravity of the solid, respectively, $\vec{V}_{solid}$ and $\theta$ are its velocity and its instantaneous angular position, respectively.

## 3 VALIDATING TEST CASES

In this section, test cases are studied in order to validate the code described here before. Following a hydrodynamic engineer point of view, only problems with rigid body and (monophasic) fluid are considered in the present paper; in such context, validation of the code focuses on the accuracy of computations, especially as far as the pressure estimation procedure is concerned[2]. Moreover, following our engineer's point of view, these test cases have been designed to be simulated on a desktop computer, in reaching a trade-off between accuracy and CPU time cost of the considered simulations. This allows evaluating the relevance and efficiency of the SPH method for pre-design studies in industrial-related issues.

### 3.1 *Validation test case #1: Simple dam break*

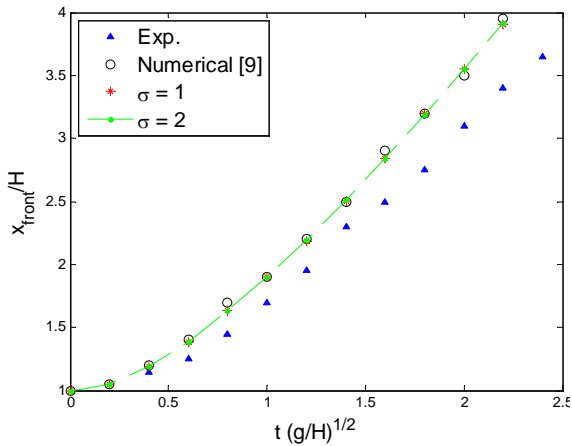

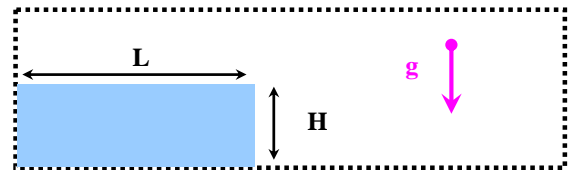

Figure 5. The "classical" dam-break problem

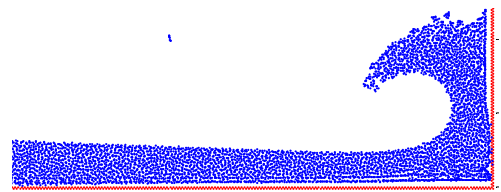

Figure 4. Time evolution of the water-front toe

Figure 6. Screenshot at $t \times \sqrt{(g/H)}=5.6$

This first simple test case reproduces one of the experiments detailed by Zhou et *al*. (1999) and also considered by Colagrossi & Landrini (2003): in this experiment, the water is initially contained inside solid boundaries of a water tank, and a piece of wax paper clamped between two metallic frames. The intense current produced by a short circuit melts the wax and quickly releases the paper diaphragm, leaving the water free to flow along a practically unlimited dry deck. For the 2D code the water is initially a square with a side-length of ~5.7 cm, and the simulation is done in using Dynamic Boundaries Conditions of SPHYSICS.

---

[2] *Problems involving elastic material, multiphase flows, etc…, which lead to high CPU time requirements, are not considered in the present paper and will be tackled in future publications.*

### 3.2 *Validation test case #2: Dam break and impact problem*

This test case is complementary to the previous one since the cinematic of the fluid is studied through a more complex case described in Colagrossi & Landrini (2003). This dam break problem with formation of spray and fragmentation reproduces the experiments in Zhou et al. (1999), where a tank of water, $H = 60$ cm and $L = 2H$, is placed at a distance $D = 3.366H$ from a vertical wall, as sketched in Fig. 5.

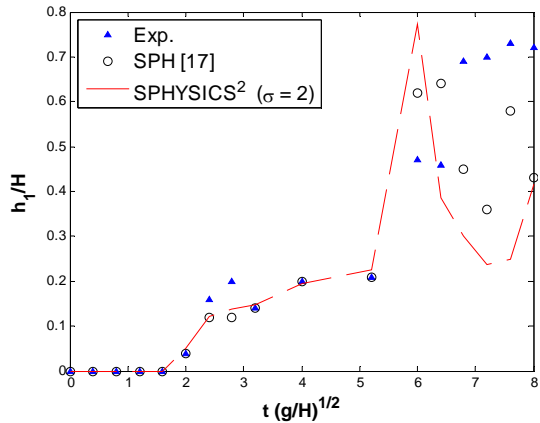 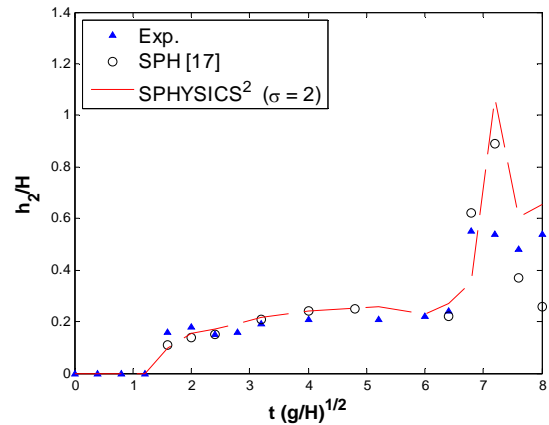

Figure 7. Total height of water at x1/H=0.825    Figure 8. Total height of water at x2/H=1.653

### 3.3 *Validation test case #3: Wedge impact in imposed motion*

This last test case is interesting because it is directly related to the slamming of a hull problem. Moreover, the local pressure estimation, and then the pressure distribution, is studied through the case. The experiment considered here has been done by Zhao et al. (1997) and compared with an SPH method by Oger et al. (2006). Besides, except for the size of the tank, the parameters chosen are the same as those of Oger et al. (2006).

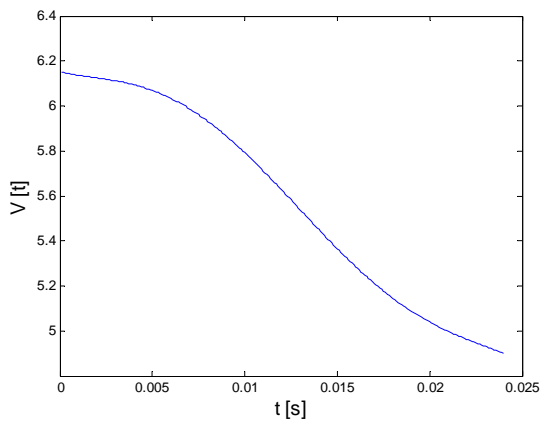 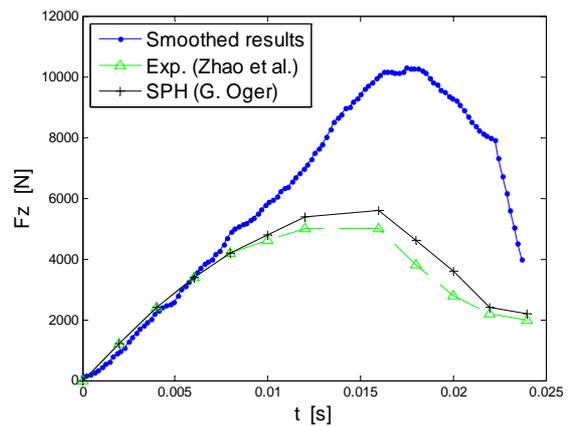

Figure 9. Experimental vertical velocity time history    Figure 10. Vertical slamming force time history

It should be mentioned that contrary to the previous cited codes, our code operates with a constant smoothing length; thus, the distance between the impact area and the tank boundary has been divided by 9 to limit CPU time, so that the side-length is now equal to 0.8 m. Because of the reflection of the sound wave generated by the impact on the tank boundaries, this means that our simulation is relevant only during the first $t = 1.6 / 80 = 0.01\,\text{s}$ (with speed of sound chosen equal to 80 m/s). After this time, there is an acoustic interaction with the wedge, which, unlike in the experiments, leads to a higher vertical force on the wedge as shown on the Fig. 10 below. Since the validity of the pressure estimation has to be confirmed, the motion of the wedge is not free but imposed by the speed recorded experimentally (as represented by Fig. 9), as described in Oger et *al.* (2006).

## 4 SOLID/SOLID CONTACTS: TOWARD COMPLEX CASES

As described in previous section, the code can simulate monophasic flow and coupling with a body in free motion can be considered. However, in some industrial-related issues, solid/solid contact has sometimes to be taken into account. Bearing this requirement in mind, additional enhancements of our code have been performed in order to tackle such situations in a simple though efficient way. These additional functionalities are described hereafter.

### 4.1 *Interaction between rigid bodies and solid boundaries.*

A lot of industrial configurations imply to manage a contact between a body and solid boundaries. A way to simulate this kind of problem is to consider the body as an elastic solid and to treat it with the SPH method as described upper. As noted in Oger et *al.* (2006), this method allows simulating accurately fluid/solid/solid interactions (Shintate & Sekine, 2004; Libersky et *al.*, Bonet et *al.*, 2004) but imposes very small time steps, which leads to rather high computational cost and consequently makes the simulation not affordable in engineering situations. Moreover, elastic materials are very sensitive to the so-called "tensile instability" (Monaghan, 2000), which can raise unexpected numerical problems. As for the fluid/solid interaction, a cheaper procedure, with rigid solid, has therefore to be developed.

A first and natural solution to manage contact is to proceed like for the fluid particles, i.e. by modifying the dynamic of the rigid body through a force which avoids interpenetration. However, this method leads to uncontrolled bounces and the contact is not physically consistent. That is why a second method, which lies on the handling of the rigid body cinematic, has been implemented in our code. Principle of the method is as follows. Noting $\vec{F}$ the result of forces on the body, $\vec{n}(n_x, n_z)$ the normal unitary vector out of the solid boundary, $\vec{t}(-n_z, n_x)$ the tangential unitary vector, $\vec{v}_1(u_1, w_1)$ and $\vec{v}_2(u_2, w_2)$ the speed of the rigid body respectively before and after the impact, we state that if $\vec{n} \cdot \vec{F} \leq 0$, then the contact is kept and it satisfies:

$$\vec{v}_2 \cdot \vec{n} = -r \cdot \vec{v}_1 \cdot \vec{n} \tag{16}$$

$$\vec{v}_2 \cdot \vec{t} = f \cdot \vec{v}_1 \cdot \vec{t} \tag{17}$$

where $r \geq 0$ is the restitution coefficient of the impact. $f$ is a friction coefficient, eventually different from unity only for an inelastic impact, i.e. $r = 0$, because then, there is no bounce and the rigid body can slide on the solid boundary. It is important to note that the proposed formulation in Eq. (15) stops the acceleration of the rigid body in case of inelastic impact. Then, in this case, $\vec{v}_1$ has to be added to the estimated effect of the acceleration $\frac{dt}{M}\vec{F}$.

The solution of the linear system (15) reads:

$$u_2 = n_z(n_z \cdot u_1 - n_x \cdot w_1) - r \cdot n_x(-n_x \cdot u_1 + n_z \cdot w_1) \tag{18}$$

$$w_2 = -n_x(n_z \cdot u_1 - n_x \cdot w_1) - r \cdot n_z(-n_x \cdot u_1 + n_z \cdot w_1) \tag{19}$$

In order to avoid any interpenetration, when the rigid body enters in the contact area, the body is lightly displaced to the borders of this area. In an intuitive way, a contact is detected when the distance between the body and the solid boundaries is smaller than the initial length between particles. This procedure is therefore very cheap in CPU time, and extents in a simple way the capacities of our code.

### 4.2 *Application to a synthetic complex case*

In what follows, the feasibility of complex simulations with both fluid/solid and solid/solid interactions is studied. Figure 11 sketches the case under study: an inelastic cylinder falls into a cylindrical tank partially filled with water. This elementary test-case, which combines various fluid/solid and solid/solid interactions, is representative of real situations encountered in safety assessment studies for instance.

A first simulation allows checking that the motion of the ball is well reproduced without water: numerical calculations are compared with the theoretical solution of the problem. The velocities calculated by the code are given on Fig. 12: the free fall of the ball can be observed twice and occurs when the horizontal velocity is constant. It is noticeable that the contact procedure introduces some numerical damping (underestimation of the maximum velocity, phase-lag between the computed and theoretical velocities). This damping is naturally not linked with the kinetic energy dissipation in the inelastic impact, given by the ratio:

$$\frac{\Delta E_c}{E_{c_1}} = 1 - \left(\frac{\vec{v}_1 \cdot \vec{t}}{v_1}\right)^2 = \sin^2(\theta) \tag{20}$$

with $\theta$ the angle between $\vec{v}_1$ and $\vec{t}$ at the impact.

Since the first oscillation of the solid in the tank is well enough predicted, a second simulation is performed to compute the solid motion with water, until $t \approx 1$ s. The cylinder considered has a mass $m = 1$ kg, which means a density $\rho \approx 466$ kg/m$^3$, and the water height is set to $h = 7$ cm. The impact of the body with the water at $t_1 \approx 0.266$ s stops its vertical motion and the body slides on the free surface with a decreasing velocity (see Fig. 13).

Then, between $t_3 \approx 0.66$ s and $t_4 \approx 0.71$ s (defined by the maximum of horizontal displacement), the rigid body slides on the tank boundaries before coming back on the free surface.

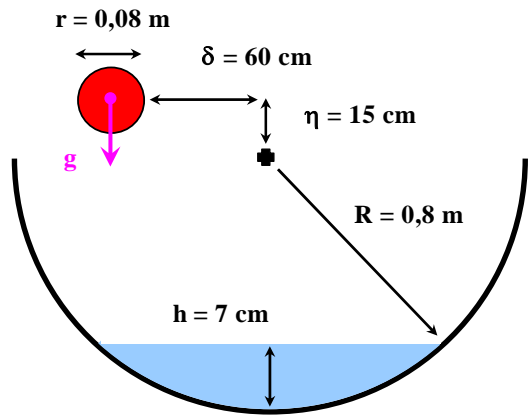
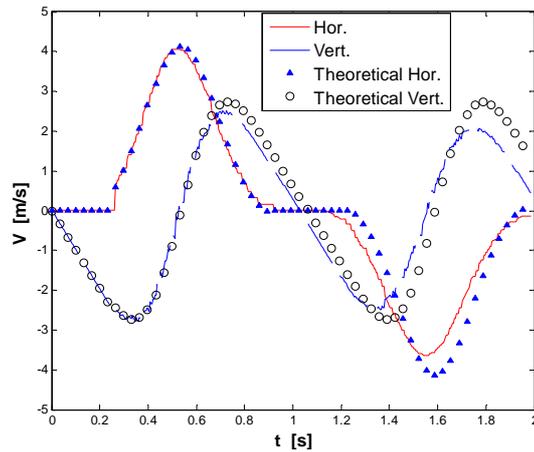

Figure 11. Sliding cylinder into a cylindrical water tank

Figure 12. Comparison between the horizontal and vertical calculated and theoretical velocities

The energy time history represented in Fig. 14 shows clearly the conservation of energy during the free fall and the kinetic energy dissipation due to the inelastic impact at time $t_1$. Figure 14 somehow confirms that the numerical damping can be neglected during the short period of contact between $t_1$ and $t_2 \approx 0.45$ s: consequently, it gives the energy lost due to the effects of water drag. The water slows down rapidly the cylinder which continues to slide at a velocity almost constant on the water. One can notice that the potential energy reference is fixed here at the rest position in the tank without water.

Although the simulation reporter here needs to be more thoroughly investigated for validation purposes, it can be concluded that the proposed developments in SPYHISC2 offer some interesting perspectives from the engineering standpoint. Complementary test cases, involving similar solid/solid and fluid/solid interactions issues are currently investigated to produce additional validation of the presented developments.

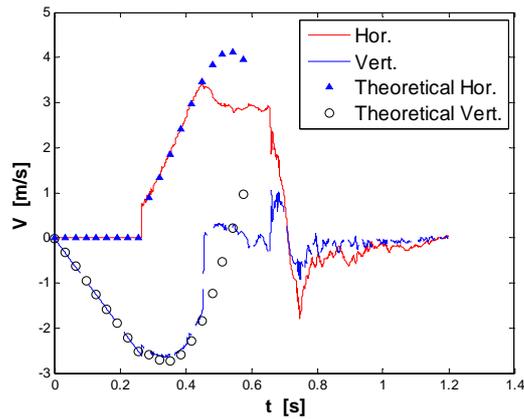

Figure 13. Comparison between the calculated velocities with water and the theoretical velocities without water

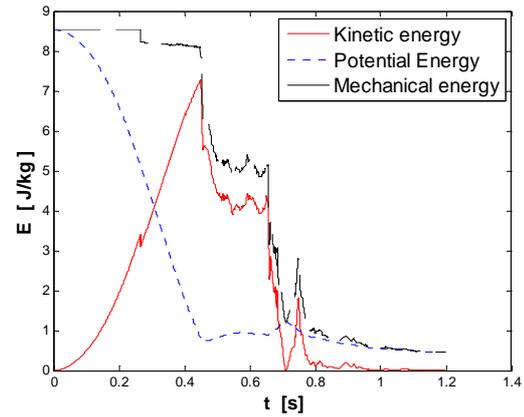

Figure 14. Energy time history (simulation)

## 5 CONCLUSIONS

A contribution to the enhancement of an existing SPH-based code, namely the SPHYSICS code, have been reported and discussed in the present paper. Our work constitutes an isolated attempt to perform some adaptation of the SPHYSIC code to enlarge the field of potential applications of the SPH technique.

Additional functionality has been implemented in the original version of the code, in order in particular to tackle solid-solid & fluid interactions. The enhanced code SPHYSICS[2] gives some perspective for future developments for engineering applications and we are currently working to secure and validate the developments presented in the paper.

Purpose of our paper is also to trigger some interest for the new functionality and we hope that reporting some of our developments will give rise to possible exchanges and collaboration on the matter.